\documentclass[a4paper,10pt]{article}

\usepackage{ucs}

\usepackage{amsmath}

\usepackage{amssymb}

\usepackage{theorem}

\usepackage{xspace}

\usepackage{fontenc}

\usepackage{graphicx}

\usepackage{ifthen}

\newcommand {\p}    {\partial}
\newcommand {\nn}   {\nonumber}

\newcommand {\arx}  [1] {{\tt#1}}
\newcommand {\cont} {\!\cdot\!}
\newcommand {\cart} {\!\wedge\!}
\newcommand {\eqn}  [1] {equation~(\ref{#1})} 
\newcommand {\teqns}[2] {equations~(\ref{#1},~\ref{#2})}
\newcommand {\Lie}  {\pounds}
\newcommand {\const}{\operatorname{const}}

\newcommand {\tdot} [1] {\stackrel{{\,}_{\scriptscriptstyle{\bullet}}}{#1}}

\newcommand {\rhs}  {r.h.s.\@\xspace}

\newcommand {\finis}{$\ \blacksquare$}

\newcommand {\Fig}  [1] {fig.~\ref{#1}}
\newcommand {\CITE} [2] {#1~\cite{#2}}

\newtheorem{lemma}       {Lemma}       [section]
\newtheorem{corollary}   {Corollary}   [section]
\newtheorem{proposition} {Proposition} [section]

\ifthenelse{\isundefined{\url}} 
{\newcommand{\tit}[1]{{\em #1} \\}}
{\newcommand{\tit}[1]{           }}

\ifthenelse{\isundefined{\url}} 
{\newcommand{\url}[1]{{\arx{#1}}}}
{}

\begin{document}

\author{Wolfgang Graf\hspace{3pt}{\thanks{email: {\tt wolfgang.graf@univie.ac.at}}}}
\title{Existence and Stability of \\
Circular Orbits in Time--Dependent \\
Spherically Symmetric Spacetimes}
\date{\today}

\maketitle

\begin{abstract}
For a general spherically four--dimensional metric the notion of ``circularity'' of a family of equatorial geodesic trajectories is defined in geometrical terms.
The main object turns out to be the angular--momentum function $J$ obeying a consistency condition involving the mean extrinsic curvature of the submanifold containing the geodesics. 
The ana\-ly\-sis of linear stability is reduced to a simple dynamical system formally describing a damped harmonic oscillator. For static metrics the existence of such geodesics is given when $J^2 > 0$, and $(J^2)' > 0$ for stability.
The formalism is then applied to the Schwarzschild--de Sitter solution, both in its static and in its time--dependent cosmological version, as well to the Kerr--de Sitter solution.
In addition we present an approximate solution to a cosmological metric containing a massive source and solving the Einstein field equation for a massless scalar.
\end{abstract}

\newpage

\section{Introduction}
\label{intro_sec}
The analysis of circular orbits (both time-- and lightlike) around an isolated massive source had always been an important tool to determine its physical parameters.
For example in Newtonian gravity, the Kepler--law $v^2=m/r$ for circular orbits in the field of a central attracting body allows to determine its mass--parameter $m$.
In the slightly generalized form $v^2=m(r)/r$ with $v^2\to\const\neq0$ it led to the assumption of dark matter in the halo of galaxies.
The same argument holds also in Einstein's gravity
for spherically--symmetric and static spacetimes.
Its description is slighty more involved, being based on an effective potential.
However at least in a cosmological context it is necessary to include the effects of expansion (possibly accelerating), thus leading to time--dependent metrics and preventing the applicability of the effective potential method.
For a possible explanation of the anomalies in the trajectories of the Pioneer probes (at approx. $20$ -- $70$ AU), such effects have been studied from diverse perspectives.
However they are far too small to be relevant in the context of the standard theory.\footnote
{for a review, see \CITE {Carrera and Giulini, 2010}{10CaG}}
In fact, a careful ana\-ly\-sis\footnote
{\CITE {Turyshev et al., 2012}{12TTK}}
of the probe geometry revealed that the acceleration anomaly can be explained away by a thermal effect.
But data concerning ``wide binaries'' ($\gg 7000$ AU)\footnote
{\CITE {Hernandez, Jimenez, Allen, 2012}{12HJA}}
seems to conform to $v^2\to\const\neq0$ and so could indicate a possible cosmological effect.

Moreover, in the last years interest has been shifted towards galactic systems with their asymptotic constant tangential velocities obeying the baryonic Tully--Fisher relation $v \to (MH)^{1/4}=\const$.\footnote
{see \CITE{McGaugh, 2005}{05Gau}}
This relation does not fit well into the conventional $\Lambda$CDM model
but is the main result of Milgrom's phenomenological MOND theory.\footnote
{for an introduction, see \CITE {Milgrom, 2008}{08Mil}}
In all these attempts the velocity of the bound orbits is of paramount importance.

Nevertheless, these studies also revealed some lack of understanding of what is meant by a ``circular orbit'' in the field of a time--dependent, but spherically--symmetric metric.
It is the aim of the present paper to propose a geometric definition of the notion of circular orbit valid in this more general setting.
This definition is substantiated by the possibility to formulate a corresponding stability--criterium.

The paper is organized as follows.
In section~\ref{conv_nots_sec} the basic notation and conventions are established.
In section~\ref{metric_sec} the equatorial metrics are defined.
In section~\ref{circ_orbs_sec} the notion of circular orbit is introduced geometrically.
In section~\ref{result_sec} the fundamental results 
for the existence and stability for circular orbits are formulated and proved.
In section~\ref{static_sec} the static case is analyzed in some detail.
In section~\ref{exmp_sec} some examples are analyzed (including time--dependent ones)
before concluding with a discussion of our results in section~\ref{disc_sec}.

The emphasis is on timelike orbits with nonvanishing angular--momentum, $J\neq0$.
However, for completeness also the case of radial timelike trajectories, $J=0$ is briefly considered, as well as lightlike trajectories.
It should be stressed that we deal exclusively with one--parameter families of trajectories sweeping the whole equatorial plane and not with isolated trajectories. 

\section{Conventions and Notations}
\label{conv_nots_sec}
Although a generalization to spherically--symmetric spacetimes with dimension $d=3$ and $d>4$ would be straightforward, we will deal exclusively with $d=4$ and Lorentz--signature $(-1,1,1,1)$. Later on we will restrict to a corresponding ``equatorial'' spacetime with $d=3$ and the induced metric. 
As our problem is fundamentally geometric, we well use extensively the conventional geometric index--free notation. As far as possible we will follow the conventions and notations of
\CITE{O'Neill, 2006}{06ONe},\footnote
{from now on, we cite this work only by its author--name} 
with the following major exceptions. Perhaps less well--known are the useful ``musical isomorphisms'' $\flat$ and $\sharp$ between vectors $V$ and one--forms $F$, $\flat: V \to V_\flat$, where $V_\flat:=g(V)$, and and its inverse
$\sharp: F \to F^\sharp$, where $F^\sharp:=g^{-1}(F)$, corresponding to the frequent ``lowering'' and ``raising'' of a simple index with the metric tensor.
This notation keeps visible the geometric origin either as a vector or as a 1--form.
In addition we will denote the contraction between a vector $V$ and a 1--form $W$ 
with the dot--operator to a scalar by $V\cont W$.\footnote
{more generally, contraction of $V$ with respect to the first (from the left) free vector slot
of the tensorial object $W$}\footnote
{the only exception to this rule is the divergence of a vector $V$, when expressed as $\nabla\cont V$} 
This spares us from the sometimes clumsy notation of O'Neill for the metric scalar--product $\langle U,V \rangle$, which now would be written either as $U\cont V_\flat$ or as $V\cont U_\flat$.
Also instead of O'Neill's notation $D_X$ for the covariant--derivative in the direction of $X$ we will use the more conventional $\nabla_X$.
However in accordance with O'Neill we will also assume a torsion--free and metric--compatible connection $\nabla_X$:
\begin{align}
& \nabla_X Y - \nabla_Y X - [X,Y] = 0, \\
& \nabla_X \langle Y,Z \rangle - \langle \nabla_X Y,Z \rangle - \langle Y,\nabla_X Z \rangle = 0,
\end{align}
valid for any vectors $X,Y,Z$.\footnote
{the only instance where we use the $\langle\:,\,\rangle$--notation}
This defines uniquely the standard Levi--Civita connection
expressed by the usual metric--based Christoffel--symbols~$\Gamma$.

We use a unit system based on powers of the light--year (ly).
Both gravitational coupling constant $\kappa := 8\pi\,G$ and velocity of light $c$ will be set to~1.

\section{Equatorial Metrics}
\label{metric_sec}
The spherical symmetry of the metric allows the line--element to be expressed in the following
\emph{canonical form} (e.g. \CITE{Carroll 2004, ch. 42}{04Car})
\begin{equation}
\label{four_met}
ds^2 = -e^{2a}\,dt^2 + e^{2b}\,dr^2 + r^2\,d\Omega^2,
\end{equation}
where $d\Omega^2$ is the standard line--element on the unit 2--dimensional sphere, $d\Omega^2=d\vartheta^2+\sin^2{\vartheta}\,d\varphi^2$,
and $a,b$ are functions only of $t,r$.
Note that in this decomposition, the radial coordinate $r$ (sometimes called ``areal radius'') is defined uniquely as the square--root of the coefficient of $d\Omega^2$.

Evidently there is the discrete reflection isometry $\vartheta\to\pi-\vartheta$.
This isometry is frequently used to motivate the restriction to the equatorial plane when considering geodesics (e.g. \CITE{Frolov and Zelnikov 2011, ch. 8}{11FrZ}).
A more refined consideration is to base it directly on the separability of the geodesic equation (e.g. \CITE{Chandrasekhar 1983, ch. 7}{83Cha}).
Here we want to indicate briefly its geometric origin by showing that the equatorial submanifold is totally geodesic.

By Lemma \ref{kb_lemma} there is the identity $K(X,Y)=-\varepsilon\,B(X,Y)$,
where $K$ denotes one--half of the Lie--derivative of the induced metric $h$ along the normalized normal vector $n:=N^\sharp$,
$B$ denotes the extrinsic curvature, and $\varepsilon:=N^\sharp\cont N = \pm1$.
Using local coordinates, with 
$e^c = r$, $N=r\,d\vartheta$, $\varepsilon=1$, we have
\begin{equation}
K_{\mu\nu} := \tfrac{1}{2}\,\Lie_n\,h_{\mu\nu} = 
\tfrac{1}{2} \left(\Lie_{e^{-c}\p_\vartheta}\,\text{diag}\,(-e^{2a}, e^{2b}, 0, e^{2c})\right)\Big|_{\vartheta=\pi/2}
= 0
\end{equation}
and so the extrinsic curvature also vanishes, $B(X,Y)=0$. 
Therefore the equatorial submanifold is \emph{totally geodesic}, implying in particular, that any geodesic in it is also a geodesic in the original manifold.\footnote
{for a more detailed formulation, see Proposition 13 of ch. 4 in O'Neill}
As we deal exclusively with geodesics in the equatorial plane, we could restrict all the following considerations to the three--dimensional
\emph{canonical equatorial metric} (i.e. the induced metric) with line element
\begin{equation}
\label{can_met}
ds^2 = -e^{2a}\,dt^2 + e^{2b}\,dr^2 + r^2\,d\varphi^2,
\end{equation}
where $a,\,b$ are functions of $(t,r)$ only.

More generally, if the dependency of a spherically symmetric metric on $\vartheta$ is only through $\sin^2\vartheta$ and $\cos^2\vartheta$, there is reflection symmetry, and again $K_{\mu\nu}=0$. This is the case for the axially--symmetric metrics with mirror--symmetry, like the Kerr--metric. Also it will be convenient to allow some extra redundancy by introducing a more general radial coordinate than the areal radius, like in most cosmological metrics.
Therefore our ana\-ly\-sis will deal more generally with the \emph{generalized equatorial metric},
with line--element
\begin{equation}
\label{gen_can_met}
ds^2 = -e^{2a}\,dt^2 + e^{2b}\,dr^2 + e^{2c}\,(d\varphi - w\,dt)^2,
\end{equation}
where besides $a$ and $b$ also $c$ and $w$ are functions only of $(t,r)$.\footnote
{$w$ is usually denoted by $\omega$}\footnote
{in fact, assuming $w=0$, the canonical equatorial metric is hard to achieve --- if at all}
Evidently there is a remaining angular symmetry given by the Killing--vector $C:=\p_\varphi$.

\section{Circular Orbits}
\label{circ_orbs_sec}
A systematic analysis of circular orbits for general spherically--symmetric time--dependent spacetimes seems not to have been done yet save for very particular cases.
Sometimes (e.g. \CITE {Sultana and Dyer, 2005}{05SuD}), the notion of ``circular orbit'' is bound to the constancy of some ``radius'' along the trajectory.\footnote
{see also \CITE {Carrera and Giulini, 2010}{10CaG}} 
For a time--dependent spacetime this can at most be achieved by some isolated orbits.
This is of course not enough to establish Kepler--like relations e.g. in the form $v=v(r)$ ($v$: tangential velocity).
Another approach is to consider ``quasi--circular orbits'' as defined by circular orbits in a static setting perturbed by cosmological dynamics (\CITE {Faraoni and Jacques, 2007}{07FaJ}).
At the other extreme is the approach of \CITE {Nolan, 2014}{14Nol} establishing the boundedness of the orbits, which asymptotically for large times approach circular orbits. However this has been shown only for some particular McVittie spacetimes. Also this would not be sufficient to obtain 
Kepler--like relations.

Here we propose a characterization of circular orbits in between these extremes, making essential use of the remaining angular symmetry provided by the Killing--vector $C:=\p_\varphi$.
Consider a congruence of non--radial trajectories (not necessarily geodesic) with tangents $T$. The symmetry is imposed by the condition $\Lie_C T=0$. These trajectories sweep out a congruence of hypersurfaces $\Sigma$, locally given by $\sigma=\const$ with $\Lie_C \sigma=0$ by symmetry. Then also $\Lie_C N=0$, where $N\sim d\sigma$ denotes the one--form normal to $\Sigma$. The hypersurface--property can be also expressed as $N\cart dN = 0$.

Our notion of ``circularity'' will be therefore defined by the following differential conditions encoding the symmetry both for $T$ as well as for $N$,
\begin{equation}
\text{\bf circdiff}
\left\{
\begin{array}{ll}
a) & [C,T]= 0, \\
b) & C\cont dN + d\,(C\cont N) = 0, \\
c) & N\cart dN = 0.
\end{array}
\right. \nn
\end{equation}

These differential conditions must be supplemented by the following algebraic conditions,
\begin{equation}
\text{\bf circalg}
\left\{
\begin{array}{ll}
a) & T\cont N = 0, \\
b) & C\cont N = 0, \\
c) & T\cont T_\flat = -1, \\
d) & N^\sharp\cont N = +1.
\end{array}
\right. \nn
\end{equation}
Here the condition a) expresses that $T$ is contained in the corresponding hypersurface defined by $N$ --- analogously condition b).
\smallskip

The angular symmetry of the equatorial metric can be expressed as\footnote
{here the corresponding coordinate--dependent expression $C_{i;k}+C_{k;i}$
looks much simpler, but lacks an immediate geometrical meaning 
as the Lie--derivative of the metric $g_{ik}$ along~$C^i$}
\begin{equation}
\text{\bf circkill}
\left\{
\begin{array}{ll}
a) & \forall X,Y: (\Lie_C\, g)(X,Y)) \equiv C\cont d(X\cont Y_\flat) + [X,C]\cont Y_\flat + [Y,C]\cont X_\flat = 0, \\
b) & C\cont C_\flat > 0,
\end{array}
\right. \nn
\end{equation}
Condition b) serves only to prevent $C$ to be a timelike Killing--vector.
\smallskip

These conditions are still independent of any geodesy of $T$, here expressed by $\nabla_T T = 0$.
To take it into account, it will be convenient to introduce also the notion of the \emph{extrinsic curvature} $B(X,Y)$ of a (nowhere lightlike\footnote
{this case was not considered by O'Neill --- there would be some technical difficulties}) hypersurface~$\Sigma$.
This object encodes the deviation between the induced connection in the hypersurface~$\Sigma$ and the connection in the ambient manifold.
In the literature several different but closely related notions are used for this --- e.g. in O'Neill, ``shape tensor'', ``shape operator'' and ``second fundamental form''.
The ``second fundamental form'' frequently also goes under the name of {extrinsic curvature}.
Following \CITE{Frankel 1998, ch. 11.4}{98Fra} we define it by means of Gauss' equation,
\begin{equation}
\label{gauss_eq}
\nabla_X Y = \bar\nabla_X Y + B(X,Y)\,N^\sharp,
\end{equation}
where $X,Y$ are any vectors in the submanifold $\Sigma$, $\bar \nabla$ denotes the induced connection and $N^\sharp$ the vector corresponding to the normalized normal one--form~$N$.
There is a useful relation to the Lie--derivative along $N^\sharp$ of the induced metric $h$ which we express in the following
\begin{lemma} {~} \\ \hspace{-5pt}
\label{kb_lemma}
For vectors $X,Y$ orthogonal to the normalized one--form $N$ normal to $\Sigma$ with induced metric $h$, there is the identity
\begin{equation}
X\cont Y\cont(\Lie_{N^\sharp} h) = - \varepsilon\,B(X,Y).
\end{equation}
{\bf Proof} \\
The proof will be based on the fundamental identity $(\nabla_X X) \cont N = \varepsilon\,B(X,X)$, which is seen to hold due to
\begin{equation}
X\cont\nabla_X N \equiv \nabla_X(X\cont N)-\nabla_X X\cont N = \varepsilon\,B(X,X)
\end{equation}
by Gauss and orthogonality.
By polarization of this quadratic relation we immediately get
\begin{equation}
(X\cont\nabla_Y + Y\cont\nabla_X)\, N = \varepsilon\,B(X,Y).
\end{equation}
Again by orthogonality this can also be written as
\begin{equation}
(X\cont Y + Y\cont X)\cont \nabla N = -\varepsilon\,B(X,Y).
\end{equation}
Here $\nabla N$ can be replaced by $\Lie_{N^\sharp} g$.
Taking instead of $g$ the induced metric $h := g -\varepsilon\,N\!\times\!N$ 
the identity still holds
and so again by orthogonality we get the claimed result. \finis
\end{lemma}
Note that for a spacelike hypersurface $\varepsilon=-1$ and so this definition agrees 
in this case with the one given e.g. in \CITE{Wald 1984, ch. 7.2}{84Wal} and in \CITE{Carroll 2004, app. D}{04Car}.

\section{Main Results}
\label{result_sec}
The above nine conditions are still independent of the geodesy of $T$, and could refer to any congruence of vectors.\footnote
{it is not claimed that they are all independent}
Note that the geodesy, $\nabla_X X=0$ (a vectorial equation), of any trajectory with tangent $X$ in $\Sigma$ with normal $N$, implies $B(X,X)=0$ (a scalar equation). 
But the converse is not true --- evidently from $B(X,X)=0$ follows only
${(\nabla_X X)\cont N}=0$.
However assuming circularity, there is a converse when taking into account angular--momentum conservation. This is expressed in proposition \ref{prop}
and its corollary.

\subsection{Timelike Orbits}
Let us first consider timelike orbits. Then the following proposition holds,

\subsubsection{Existence}
\begin{proposition} {~} \\ \hspace{-5pt}
\label{prop}
Assuming the conditions {\bf circdiff, circalg} and {\bf circkill}, then
the necessary and sufficient conditions for geodesy, ${\nabla_T T=0}$,
are ${B(T,T)=0}$ together with angular--momentum conservation $T \cont d(C\cont T_\flat)=0$.
\smallskip

\noindent
{\bf Proof} 
\begin{enumerate}
\item {\em necessity.} 
Here we assume geodesy $\nabla_T T=0$. 
\begin{itemize}
\item [a)] Differentiating assumption {\bf circalg a)} with $\nabla_T$ and using geodesy gives 
$1/2\,(\Lie_{N^\sharp}g)(T,T))=0$. By Lemma \ref{kb_lemma} this is equivalent to $B(T,T)=0$, as claimed. 
\item [b)] Contracting assumption {\bf circdiff a)} with $T_\flat$ and using assumption {\bf circalg c)} as well as the geodesy, results in $T\cont d(T\cont C_\flat)=0$, as claimed.
\end{itemize}
\item {\em sufficiency.}
Here we assume $B(T,T)=0$ and $T\cont d(T\cont C_\flat)=0$. 
\begin{itemize}
\item [a)] Contracting {\bf circdiff a)} with $T_\flat$ and using $T\cont d(T\cont C_\flat)=0$
gives ${\nabla_T T\cont C_\flat=0}$. 
\item [b)] Applying $\nabla_T$ on {\bf circalg a)} and using ${B(T,T)=0}$ in its equivalent form  ${K(T,T)=0}$ gives ${\nabla_T T\cont N=0}$. 
\item [c)] Applying $\nabla_T$ on {\bf circalg c)} results in ${\nabla_T T\cont T_\flat=0}$. 
\end{itemize}
As $(C_\flat,\,N,\,T_\flat)$ constitutes a complete system of linearly independent~1--forms, there follows geodesy $\nabla_T T=0$, as claimed. \finis
\end{enumerate}
\end{proposition}
Disregarding the algebraic conditions, the conditions 
of the previous proposition constitute a simultaneous differential system both for $T$ and implicitly for $N$.
Defining the angular--momentum of $T$ by $J:=C\cont T_\flat$ and assuming $J\neq0$,
this can be made more explicit by the following
\begin{corollary} {~} \\ \hspace{-5pt}
\label{coro}
For the generalized equatorial metric of \eqn{gen_can_met} define the following vectors 
$C$, $T$, one--form $N$ and scalar $\sigma\,$\footnote
{we fix some arbitrary signs by requiring $P>0$, $Q>0$}
\begin{align}
C := &\ \p_\varphi, \\
\label{n_def_J}
N := &\ \frac{1}{\sigma}\,(\dot J\,dt + J'\,dr), \\
\label{t_def_J}
T := &\ \frac{P}{Q}\left[J'\,\p_t - \dot J\,\p_r + w\,J'\,\p_\varphi \right] + J\,e^{-2c}\,\p_\varphi, \\
\label{s_def_J}
\sigma := &\ Q\,e^{-(a+b)}, \\
& \text{where} \quad P^2 := 1 + J^2\,e^{-2c}, \quad Q^2 := e^{2a}J'^2 - e^{2b}\dot J^2. \nn
\end{align}
Then all the assumptions {\bf circdiff, circalg} and {\bf circkill} of Proposition \ref{prop} are satisfied, including angular--momentum--conservation.
And the necessary and sufficient condition for geodesy can be expressed as the single 
{\em J--equation}
\begin{equation}
\label{j_eqn_J}
K = -\left(\frac{1+2\,J^2/e^{2c}}{P^2\,Q^2}\left(e^{2a}\,J'c' - e^{2b}\,\dot J\dot c \right) 
- \frac{J\,w'}{PQ}\right)\sigma,
\end{equation}
where $K := -\nabla\cont N^\sharp$ is the {\em mean curvature} of $\Sigma$,\footnote
{taking into account the signature of the metric,
we define it in general as $K := -\varepsilon\,\nabla\cont N^\sharp$} 
explicitly given by
\begin{align}
K = &\ \frac{1}{\sigma\,Q^{2}}\left[ \left( J'^2\,\ddot J + \dot J^2\,J'' - 2\,J'\dot J\,\dot J' \right)  
- e^{2(a-b)}(a'+c')\,J'^3 - e^{2(b-a)}(\dot b+\dot c)\,\dot J^3 \right. \nn \\
&\ \left. \qquad +~(2a'-b'+c')\,\dot J^2 J' + (2\dot b - \dot a + \dot c)\,J'^2\dot J \, \right].
\end{align}

\noindent
{\bf Proof} \\
All the assumptions of Proposition \ref{prop} are evidently satisfied.
Also angular--momentum conservation is seen to hold due to $\nabla_T J \equiv T\cont dJ = T\cont(\sigma\,N)=0$.
By Lemma \ref{kb_lemma} and orthogonality ${T\cont N=0}$,
\begin{equation}
B(T,T) = 0 \quad \Longleftrightarrow \quad T\cont T\cont (\Lie_{N^\sharp} g) = 0.
\end{equation}
The inverse metric
\begin{equation}
g^{-1} = -e^{-2a}(\p_t + w\,\p_\varphi)^2 + e^{-2b}\,\p_r^2 + e^{-2c}\,\p_\varphi^2
\end{equation}
can also be expressed by the quasi--orthogonal decomposition
\begin{equation}
g^{-1} = P^{-2}\left( -T\!\times\! T + C\!\times\! C\,e^{-2c} - J\,(C\!\times\! T + T\!\times\! C)\,e^{-2c} \right) + N^\sharp\!\times\! N^\sharp.
\end{equation}
Therefore we can write
\begin{align}
P^{-2} T\cont T\cont (\Lie_{N^\sharp} g) \equiv &\ -\nabla\cont N^\sharp + N^\sharp\cont N^\sharp\cont (\nabla N) \nn \\
+ &\ P^{-2} e^{-2c}\left(C\cont C\cont (\nabla N) - J\,(C\cont T\cont + T\cont C\, \cont\,) (\nabla N)\right).
\end{align}
Evaluating the terms on the \rhs, then
$N^\sharp\cont N^\sharp\cont (\nabla N) = 0$ by normalization, 
${C\cont C\cont (\nabla N) =} \tfrac{1}{2} N^\sharp\cont d(e^{-2c})$ by orthogonality and symmetry, and \\
${(C\cont T\cont + T\cont C\, \cont\,) (\nabla N)} = N^\sharp\cont T\cont d((d\varphi-w\,dt)\,e^{2c})$, again by orthogonality and symmetry,
so that finally,
\begin{equation}
P^{-2}B(T,T) \equiv K + \left(\frac{1+2\,J^2/e^{2c}}{P^2\,Q^2}\left(e^{2a}\,J'c' - e^{2b}\,\dot J\dot c  \right) - \frac{J\,w'}{PQ}\right)\sigma = 0,
\end{equation}
as was to be demonstrated. \finis
\end{corollary}
The case of vanishing angular--momentum, $J=0$, can be dealt with similarly.
For a hypersurface $\Sigma$, locally defined by $S=\const$,
define the vector $T$, one--form $N$ and scalar $\sigma$ now as
\begin{align}
\label{n_def_J0}
N := &\ \frac{1}{\sigma}\,(\dot S\,dt + S'\,dr), \\
\label{t_def_J0}
T := &\ \frac{1}{Q}\left[S'\,\p_t - \dot S\,\p_r + w\,S'\,\p_\varphi \right], \\
\label{s_def_J0}
\sigma := &\ Q\,e^{-(a+b)}, \\
& \text{where} \quad Q^2 := e^{2a}S'^2 - e^{2b}\dot S^2. \nn
\end{align}
Then again all the assumptions {\bf circdiff, circalg} and {\bf circkill} of Proposition \ref{prop} are satisfied, including angular--momentum--conservation in the form $J=0$.
The necessary and sufficient condition for geodesy can then be expressed as the single 
{\em S--equation}
\begin{equation}
\label{j_eqn_J0}
K = -\left(e^{2a}\,S'c' - e^{2b}\,\dot S\dot c  \right) \frac{\sigma}{Q^2},
\end{equation}
where $K := -\nabla\cont N^\sharp$ again denotes the {\em mean curvature} of $\Sigma$.
\medskip

So now we are left with only one nonlinear second--order partial differential equation for one unknown $J(t,r)$ (resp. $S(t,r)$), albeit with such a high complexity (comparable to the closely related and notoriously complex equation for mi\-ni\-mal surfaces), that in a non--stationary setting only in very special cases it can be hoped to get an exact solution. 
A corresponding circular orbit is said to exist in a certain $(t,r)$--region, if the solution $J$ (resp. $S$) is nonvanishing and real there, so that $J^2>0$ (resp. $S^2>0$).
Once we have such a solution, any other relevant quantity can be derived from it --- in particular, the tangents $T$ to the geodesics from \eqn{t_def_J} (resp. \eqn{t_def_J0}).
For the canonical equatorial metric given by \eqn{can_met}, 
if $J'\neq0$ the equation $J(t,r) = j_0 = \const$ can be solved implicitly to give the time--development of the areal radius, $r = f(t;j_0)$ (similarly for $S$).

\subsubsection{Stability}
\label{stab_sec}
Here we will analyze the stability of the solution of the J--equation for timelike orbits under \emph{li\-near perturbations}. For this purpose, consider the perturbed geodesic with normalized tangent $\tilde T:=U-T$ and perturbation $U$.
Then from geodesy and normalization of $T$ in first--order we must have
\begin{equation}
U\cont T_\flat = 0, \quad \nabla_{T} U = -\nabla_{U} T.
\end{equation}
Note that in order to have a more conventional form for the resulting dynamical system,
we will now use the thick--dot notation $\tdot \lambda$ instead of $\nabla_T \lambda$.\footnote
{not to confound with the thin--dot notation $\dot \lambda \equiv \p\lambda/\p t$}
We must also distinguish the general case $J\neq0$ from the special case $J=0$.
\begin{proposition} {~} \\ \hspace{-5pt}
Define $\mu := U\cont C_\flat$ and $\nu := U\cont N$ and assume nonradial trajectories,
$J\neq0$.
Then there results the system
\begin{align}
\label{ds_mu}
\tdot \mu &\ = -\sigma\,\nu, \\
\label{ds_nu}
\tdot \nu &\ =  \quad\! \kappa\,\mu + \delta\,\nu, 
\end{align}
\begin{equation}
\text{where} \quad \kappa := 2\,\frac{\sigma\,J}{P^2Q^2}\,(e^{2a}c'J' - e^{2b}\dot c \dot J)\,e^{-2c} 
 + \frac{\sigma}{PQ}\,w', \quad \delta := {\tdot \sigma}/\sigma. \nn
\end{equation}
{\bf Proof} \\
Condition $U\cont T_\flat=0$ with $U\cont C_\flat = \mu$ and $U\cont N = \nu$ is solved with
\begin{equation}
U := \mu\,\frac{e^{-2c}}{P^2}\,(C + J\,T) + \nu\,N^\sharp.
\end{equation}
Decomposing $\nabla_{T} U = -\nabla_{U} T$ 
into $\tdot \mu = -\nabla_U J$ and $\tdot \nu = 2\,\text{symm}(\nabla N)(U,T)$; inserting $U$
and using the angular symmetry as well as the J--equation gives the above system of equations.
\finis
\end{proposition}
The special case $J=0$ can be solved similarly by setting
\begin{equation}
U := \mu\,e^{-2c}\,C + \nu\,N^\sharp,
\end{equation}
with
\begin{align}
N := &\ \frac{1}{\sigma}\,(\dot S\,dt + S'\,dr), \\
\sigma := &\ Q\,e^{-(a+b)}, \\
& \text{where} \quad Q^2 := e^{2a}S'^2 - e^{2b}\dot S^2, \nn
\end{align}
resulting in the following coefficients for the system of \teqns{ds_mu}{ds_nu},
\begin{equation}
\kappa := \frac{\sigma}{Q}\,w', \quad \delta := {\tdot \sigma}/\sigma. \nn
\end{equation}
In the case purely spherically--symmetric case $w=0$,
the system can be immediately integrated to
$\nu = c\,\sigma$, $\mu = -c\,\sigma^2 + k$, with constants $c,\,k$.
For such radial timelike geodesics a boundedness criterium does not make much sense.
Here we propose a criterium based on the expansion of the trajectory: the additional expansion $c\,\nabla\cont(\sigma N^\sharp)$ should remain smaller than the unperturbed expansion $\Theta := \nabla\cont T$. This amounts to the condition
\begin{equation}
|c|\,|\Delta\, S| \ll |\Theta|,
\end{equation}
\medskip
where $\Delta$ denotes the d'Alembertian operator.\footnote
{note the relation 
$\Delta\, S = -\varepsilon\,\sigma K + N^\sharp\cont\nabla \sigma$}

For the canonical equatorial metric (i.e. $c=1/2\,\ln r^2$, $w=0$) and $J\neq0$ the coefficients of dynamical system simplify somewhat.
Formally an angular frequency $\omega$ for the perturbation can be introduced,
\begin{equation}
\label{can_om}
\omega^2 := \kappa\,\sigma \equiv \frac{1}{r}\,\frac{(J^2)'}{r^2+J^2},
\end{equation}
as well as a formal damping--term\footnote
{note that for $\delta$ to be well--defined, we must have $J'\neq0$}
\begin{equation}
\delta := {\tdot \sigma}/\sigma, \quad \text{where} \quad \sigma \equiv 
J'\, e^{-b} \left(1 - Y^2\right)^{1/2}, \quad \text{with} \quad  Y := e^{b-a}\,\dot J/J'.
\end{equation}
These expressions are valid even in the time--dependent case.
Whereas $\kappa$ ($=\omega^2/\sigma$), $\sigma$ and~$\delta$ do also depend on the metric, $\omega$ does not.
\smallskip

The above system of equations \teqns{ds_mu}{ds_nu} constitutes a system of first--order linear--homo\-ge\-neous 
ordinary differential equations $\tdot x = A(\tau)\, x$ for $x := (\mu,\nu)^T$ 
with time--dependent coefficient--matrix $A$.
It has the form of the equation for a \emph{damped harmonic oscillator}.
Assuming the trajectories to be timelike geodesically complete,
it can be considered as a \emph{nonautonomous dynamical system}.
The time--dependency comes from functions 
explicitly dependent on the affine parameter through the metric.
The analysis of the stability of our circular orbits is thus reduced
to the analysis of the stability of a linear nonautonomous dynamical system.
Unfortunately the results of the stability theory for autonomous dynamical systems do not carry over.
The appropriate mathematical notions are not as easy to apply and are outside the scope of this paper.\footnote
{there is already an extensive literature on nonautonomous dynamical systems--- see e.g. \CITE{Kloeden and P\"otzsche, 2013}{13KlP} for an introduction}
Only in the static case it could be considered an autonomous dynamical system,
with associated two--dimensional phase--space, where the stability--behaviour is well--known.\footnote
{see section \ref{static_sec} on static spacetimes}

\subsection{Lightlike Orbits}
The lightlike case is somewhat special and cannot immediately be dealt with by adapting the assumptions leading to proposition \ref{prop}. Instead, we follow a more direct way.

\subsubsection{Existence}
It is well--known (e.g. \CITE{Wald 1984, ch. 4.2}{84Wal}) that the normal vector to a null--hypersurface is automatically geodesic. That is, assuming the null--hypersurface $\Sigma$ locally defined by $S=0$ with $(dS)^\sharp\cont dS=0$, then $T := N^\sharp$ with $N:=dS$ is an affinely parametrized null--geodesic, $\nabla_T T=0$, $T\cont T_\flat=0$. This holds a fortiori also for a congruence $\Sigma_\lambda$ of such hypersurfaces, locally defined by $S=\lambda=\const$.

Now, consider a symmetry of the metric given by the Killing--vector $C$ and impose it also on the one--form $N$, by requiring 
\begin{equation}
\Lie_C N \equiv d(C\cont N) + C\cont (dN) =0.
\end{equation}
The last term vanishes due to $N=dS$, and we are left with the condition $C\cont N=\const$.\footnote
{e.g. the Schwarzschild line--element can be written
$ds^2 = X\,(dt + 1/X\,dr)\,(dt - 1/X\,dr) + r^2 d\Omega^2$, where $X := 1 - 2\,m/r$.
With $N_\pm := dt \pm 1/X\,dr$ we have $N_\pm^2=0$ and $N_\pm = d\,S$ with $S := t \pm \int 1/X\,dr$.
For the timelike Killing--vector $C:=\p_t$, in fact $C\cont N=1=\const$}
This is somewhat stronger than the usual conservation along $N^\sharp$, $N^\sharp\cont d(C\cont N)=0$, which we would have obtained from $\Lie_C N^\sharp \equiv [C,N^\sharp]=0$ by contraction with $N$ and geodesy.
However, taking the angular symmetry given by $C=\p_\varphi$, then we automatically have vanishing angular--momentum $J := C\cont N = 0$. 
This is immediately evident by using the canonical double--null form as given by
\CITE {Hayward, 1996}{96Hay},
\begin{equation}
ds^2 = -2\,e^{-f}du\,dv + R^2 d\Omega^2,
\end{equation}
where $f,R$ depend only on $u,v$. This form of the metric is unique up to coordinate transformations $u=\tilde u(u),\:v=\tilde v(v)$. Its ``equatorial'' form is obtained again by setting $\vartheta=\pi/2$ in the $S_2$--part of the metric.
Without loss of generality we can set $N:=du$. Evidently $C\cont N \equiv 0$, so that the only null geodesics compatible with spherical symmetry are the radial null geodesics, with $J=0$.
For the existence, this is the only constraint.

\subsubsection{Stability}
Again, for linear stability for $\nabla_{N^\sharp} N=0$ with $N^\sharp\cont N=0$ we must require $U\cont N=0$ and $\nabla_{N^\sharp} U = -\nabla_{U} N^\sharp$.
Introducing the auxiliary null vector $M$, $M\cont M_\flat=0$, with $M\cont N=-1$
and $M\cont C_\flat=0$, the algebraic condition can be satisfied with
\begin{equation}
U = \mu\,e^{-2c} C - \nu\,N^\sharp,
\end{equation}
so that $\mu = U\cont C_\flat$ and $\nu = U\cont M_\flat$.
Using geodesy and $C$--symmetry, we arrive at the trivial system
\begin{align}
\label{ds_mu_null}
\tdot \mu &\ = 0, \\
\label{ds_nu_null}
\tdot \nu &\ = 0. 
\end{align}
Therefore linear stability does not require any constraint in addition to $J=0$.
The condition $\nu=\const$ amounts to a constant rescaling of $N$, whereas $\mu=\const$ leads to a constant rotation of the trajectories and so in this sense they are rigid.

\section{Static Spacetimes}
\label{static_sec}
Here we consider explicitly static metrics in the sense of $\dot a = \dot b = \dot c = 0$ and $w=0$ in the region of interest. Although in this case our algorithm is (as of course it should) completely equivalent to the very well--known algorithm based on the effective potential,
it is more directly applicable. For example, in this case the fundamental J--equation of \eqn{j_eqn_J} is immediately solved by the simple algebraic relation
\begin{equation}
\label{j2_eqn}
J^2_{\scriptscriptstyle s\!t\!a\!t} = \frac{a'}{c'-a'}\,e^{2c}.
\end{equation}
Assuming $e^{2c}>0$ the necessary and sufficient condition for $J^2 >0$ is
either 
\begin{equation}
2\,a' > c' > a' > 0 \quad \text{or} \quad 2\,a' < c' < a' < 0.
\end{equation}
The local tangential velocity with respect to the static observers
$U = e^{-a}\p_t$,
\begin{equation}
\label{v2_stat}
v^2_{\scriptscriptstyle s\!t\!a\!t} := -\frac{g_{rr}\tdot r^2 + g_{\varphi \varphi}\tdot\varphi^2}{g_{tt}\tdot t^2} \equiv
\frac{1+g_{tt}\tdot t^2}{g_{tt}\tdot t^2} = \frac{J^2\,e^{-2c}}{1+J^2\,e^{-2c}} = \frac{a'}{c'}.
\end{equation}
Note that as long as $U$ remains timelike and $J^2>0$, then $0 < v^2 < 1$.
\smallskip

In particular, for the canonical equatorial metric, this reduces to the extremely simple relation $v^2_{\scriptscriptstyle s\!t\!a\!t} = r\,a'$.
Let us also note that in view of the asymptotically constant galactic rotation curves,
requiring constant velocity $v$ immediately results in the well--known relation\footnote
{e.g. \CITE{Roberts, 2002}{04Rob}} $a = {v^2}\,\ln{(r/r_0)}$.
This is not very satisfying, as the corresponding metric is not asymptotically Lorentzian.
\smallskip

The stability analysis also becomes particularly simple, as the dynamical system becomes undamped autonomous, with angular frequency\footnote
{see also \eqn{can_om} for a simpler expression using the canonical equatorial metric}
\begin{equation}
\label{ang_freq}
\omega^2 := \kappa\,\sigma = (1+J^2\,e^{-2c})^{-1}\,(J^2)'\,c'\,e^{-2(b+c)}.
\end{equation}
Assuming $c'>0$ in addition to $J^2>0$, the condition for stability reduces to the positivity of the squared frequency,
\begin{equation}
\omega^2\sim (J^2)' \sim a''/a'-2\,a' - (c''/c'-2\,c') > 0.
\end{equation}
For the canonical equatorial metric, the existence criterium $J^2>0$ then reduces to $1>r\,a'>0$, 
whereas the stability $(J^2)'>0$ criterium reduces to
\begin{equation}
r^2\,a'' + 3\,r\,a' -2\,(r\,a')^2 > 0.
\end{equation}
Both these criteria for circular orbits in a static spherically--symmetric metric can be found e.g. in \CITE{Lake, 2004}{04Lak}.

Recapitulating, the conditions for existence and stability of circular orbits in a static spherically symmetric metric are just the two conditions $J^2>0$ and $(J^2)'>0$, where $J^2$ is defined by \eqn{j2_eqn}. Energy conservation is not invoked at all.
This sets our algorithm apart from the conventional one where an effective potential $V(r;L,E)$ has first to be set up, and where the constant parameters $L,E$ have to be chosen so that for a particular~$r$, the three conditions $V=0$, $V'=0$ and ${V''>0}$ are satisfied. Also the direct physical interpretability of $J$ vs. $V$ has the advantage of a better contextuality ---
after all we are dealing primarily with a problem having angular symmetry.
But the most salient advantage is its applicability to a time--dependent setting, where the conventional approach inextricably based on energy conservation cannot be generalized.

\section{Examples}
\label{exmp_sec}
In the following we will first apply our algorithm to some spacetimes which are better known under the generic name {\em Schwarzschild--de Sitter spacetimes}. They correspond to metrics solving the Einstein--equation including a de Sitter--term, $\mathit{Ein} = -\Lambda\,g$ ({\em Ein} denotes the Einstein--tensor) and reducing to the Schwarzschild--solution for $\Lambda=0$. Although they are (for $\Lambda\geq0$) locally isometric, their metric tensors differ markedly.
To keep these particular solutions apart, we will here refer them by the name of their discoverers appended to the name of Schwarzschild: \\
i)~~{\em Schwarzschild--Kottler metric} (\CITE{Kottler, 1918}{18Kot}) and \\
ii)~{\em Schwarzschild--Robertson metric}
(\CITE{Robertson, 1927}{27Rob}).
\smallskip

\noindent
We will apply our algorithm also to the
{\em Kerr--de Sitter metric}, which generalizes the Schwarzschild--Kottler metric
to a stationary rotating spacetime. 
\smallskip

\noindent
As an additional explicitly time--dependent example we will consider a particular {\em cosmological solution with a scalar field and massive source}, where the J--equation can be solved approximatively.

\subsection{Schwarzschild--Kottler Metric}
Here
\begin{equation}
\label{sds_stat}
e^{2a} = e^{-2b} := 1 - 2\,\frac {m}{r} - \frac {1}{3}\,\Lambda\,r^2, \quad e^{2c} = r^2,
\end{equation}
so that $a+b=0$.\footnote
{we will not write out explicitly the $e^a$--terms in the following}
In view of its cosmological reformulation in the next section, we assume $K^2 := \frac{1}{3}\,\Lambda\,m^2\geq0$. 
In addition assuming $K^{2} < 1/27$, the equation $e^{2a}=0$ has two positive real solutions $r_\pm$, with $e^{2a}>0$ in the ``static'' range $r_- < r < r_+$.
$r_b :=r_- > 2\,m$ denotes the black hole horizon, whereas $r_s :=r_+ < 1/H$, where $H^2:=\frac {1}{3}\,\Lambda$, denotes the static limit.\footnote
{for a more detailed discussion of the global structure, see e.g. \CITE{Beig and Heinzle, 2005}{05BeH}}
Despite the cosmological $\Lambda$--term this metric is evidently static in the above range, with Killing--vector $\p_t$.
Stable circular orbits in such metrics have already been studied by \CITE{Stuchl\'ik and Hled\'ik, 1999}{99StH} and more recently by \CITE {Nolan, 2014}{14Nol} in the context of 
(non--static) McVittie metrics.

The equatorial circular trajectories satisfy the J--equation in the purely algebraic form of \eqn{j2_eqn}. Defining the \emph{reduced areal radius} $\varrho := r/m$,
\begin{equation}
\label{jj_ksds}
J^2_{\scriptscriptstyle S\!K} = m^2\varrho^2\,\frac{1-K^2 \varrho^3}{\varrho-3},
\end{equation}
which agrees with the conventional calculation\footnote
{we will also not write out the explicit form of the $J$--terms in the following}
and is positive in the range ${3 < \varrho < K^{-2/3}}$.
This existence range slightly overlaps the above static range, the higher bound exceeding the static limit.
Without solving any differential equation, the corresponding trajectories can now be calculated directly from this $J$ using \eqn{t_def_J},
\begin{equation}
T_{\scriptscriptstyle S\!K} = \left( 1 + \frac {J^2}{r^2}\, \right)^{1/2} e^{-a} \, \p_t + \frac {J}{r^2}\,\p_\varphi.
\end{equation}
Then the corresponding local tangential velocity with respect to static observers defined by 
$U := e^{-a} \p_t$ is simply\footnote
{see \CITE{Abramowicz, 2016}{16Abr}}
\begin{align}
\label{v2_sds}
v^2_{\scriptscriptstyle S\!K} := &\ \frac{e^{2b}\tdot r^2 + r^2\tdot\varphi^2}{e^{2a}\tdot t^2}
\equiv \frac{e^{2a}\tdot t^2-1}{e^{2a}\tdot t^2} \\
= &\ J^2/(r^2 + J^2) = \frac{1-K^2\varrho^3}{\varrho-K^2\varrho^3-2},
\end{align}
which is evidently positive in the range where $J^2$ is and vanishes for $\varrho=K^{-2/3}$. 
In this range it is always smaller than the ``bare'' velocity ${v^2_0 := 1/(\varrho-2)}$ ---
a manifestation of the repulsive character of the $\Lambda$--term.
More precisely,
\begin{align}
\label{v2_sk}
v^2_{\scriptscriptstyle S\!K} = &\ \frac{1-K^2\varrho^3}{\varrho-2-K^2\varrho^3} 
\equiv \frac{1}{\varrho-2} - \frac{\varrho-3}{(\varrho-2)^2}
\left(1 - \frac{K^2\varrho^3}{\varrho-2}  \right)^{-1} K^2\varrho^3 \nn \\
\approx &\ \frac{1}{\varrho-2}\left(1 - \frac{\varrho-3}{\varrho-2}\,K^2\varrho^3\right)
\end{align}
up to $o(K^2\varrho^3)^2$,
correcting the Kepler--relation $v^2=1/\varrho$ for $\varrho\gg1$ as long as $K^2\varrho^3\ll1$.

Let us make some rough order of magnitude estimates.
Assuming a current Hubble--parameter $H=75$ km/s\,Kpc $\approx7.6\times10^{-11}\:ly^{-1}$,\footnote
{we will use throughout units of lightyears (ly)} 
then for a mass of the solar--system ($m\approx1.6\times10^{-13}\:ly$) and the orbit of Neptune ($r \approx30\: AU\approx 4.8\times10^{-4}\:ly$), we have $K\approx 1.2\times 10^{-23}$ and $\varrho \approx 3.1\times 10^9$.
The correction--factor $\Delta$ (where $v_s := v_0\,(1+\Delta)$) of the bare Kepler--relation 
results as $\Delta = -1/2\,K^2\varrho^3 \approx -2.0\times10^{-18}$ and so would be negligible. 
\smallskip

However for galaxies like the Andromeda--galaxy, with $m_G \approx 1.5\times10^{11}\:m_\odot\approx2.3\times10^{-2}\:ly$\footnote
{without the mass of the ``dark matter halo''} 
the situation is more favourable.
For a circular orbit at the visible rim with $r\approx1.1\times10^5 \:ly$, 
$K\approx 1.8\times 10^{-12}$ and $\varrho \approx 4.7\times10^6$, giving
$\Delta \approx -1.6\times10^{-4}$, which is still below current empirical verification.
However, extending beyond the luminous region into the ``dark matter halo'' --- e.g. taking $18\,\varrho$ would already give vanishing velocity.
So, if the asymptotically flat galactic rotation curves 
admit an explanation within the current $\Lambda$CDM--paradigm, this negative effect of the repulsive $\Lambda$--term should be taken into account.
\smallskip

The squared (unperturbed) angular frequency can be defined as usual by $\Omega^2 := J^2/r^4$, giving
\begin{equation}
\Omega^2 = \frac{1}{m^2}\, \frac{1-K^2\varrho^3}{(\varrho-3)\,\varrho^2}.
\end{equation}
The equations for the dynamical system (\ref{ds_mu}, \ref{ds_nu}) now have constant coefficients and no damping, $\delta=0$, so their solutions are periodic in proper time, with angular frequency by \eqn{ang_freq}
\begin{equation}
\omega^2 = \frac{1}{m^2}\,\frac{\varrho-6 + K^2\varrho^3\,(15-4\,\varrho)}{(\varrho-3)\,\varrho^3}.
\end{equation}

Therefore the perturbation is bounded and must be considered as stable.
The assumption of geodesic completeness for the system to be considered a proper DS is trivially satisfied in this case.

The perturbed angular frequency allows to calculate the perihelion precession 
as e.g. \CITE{Wald 1984, ch. 63}{84Wal}, giving this time a correction--factor 
$\Delta = 1/2\,K^2\varrho^4$.\footnote
{neglecting eccentricity, this is in accordance with \CITE {Kerr, Hauck and Mashhoon, 2003}{03KHM}}
For the Andromeda galaxy and a mass point at the rim, this would result in enhancing the standard precession--rate of $4.2\times10^{-15}$ rad/y of the corresponding basic angular velocity of $\Omega\approx4.7\times10^{-9}$ rad/y by a factor of 250 --- far to small to be empirically relevant.
\smallskip

\begin{figure}[h t]
\includegraphics[width=12cm]{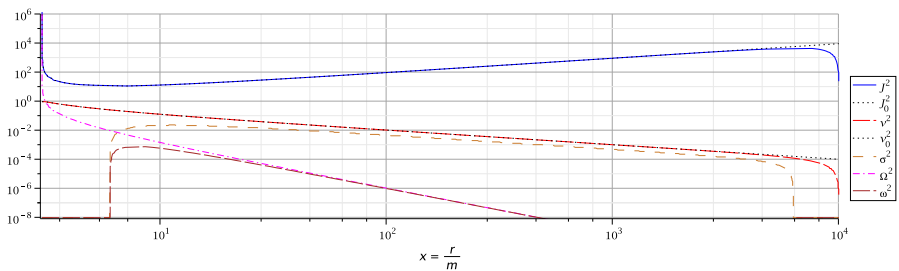}
\caption{SdS--LogLog--curves for $J^2,\,J_0^2,\,v^2,\,v_0^2,\,\sigma^2,\,\Omega^2,\,\omega^2$  ($K=1.0\times10^{-6}$)}
\label{SdS_new_loglogplots}
\end{figure}

In the following \Fig{SdS_new_loglogplots} the relevant quantities for $K=1.0\times10^{-6}$ as well as some of their reference quantities for $K=0$ are displayed.\footnote
{for galaxies $K\approx 10^{-12}$ would be more realistic,
but the plots would be less compelling}
The plot of $\sigma^2\sim (J^2)'$ nicely illustrates (better than $\omega^2$) the stability--range $(J^2)'>0$, properly contained in the existence--range $J^2>0$.
This range is given by ${6 < \varrho_{min} < \varrho < \varrho_{max} < (2\,K)^{-2/3}}$, 
where the exact limits could be obtained from a quartic equation.

\subsection{Schwarzschild--Robertson Metric}
A closely related metric is given by the line--element
\begin{align}
\label{sds_cosm}
ds^2 = &\ -\left( \frac{1-X}{1+X} \right)^2 dt^2 + e^{2H\,t}\,(1+X)^4 \left( dr^2 + r^2 d\Omega^2 \right), \\
&\ X := \frac{m}{2\,r}\,e^{-H\,t}, \quad H := K/m \equiv \sqrt{\Lambda/3}=\const, \nn
\end{align}
which is regular for $X\neq\pm1$ and $r\neq0$.\footnote
{all the second--order invariants are regular for $X\neq-1$}
In fact, as already shown by \CITE{Robertson, 1928}{28Rob}, for $\Lambda>0$ this explicitly time--dependent metric is locally isometric to the Schwarzschild--Kottler metric defined by \eqn{sds_stat}, transforming it by means of the coordinate--transformation
\begin{align}
\label{rob_t}
t \to &\ t + F\left( \frac{1-X}{1+X} \right), \\
\label{rob_r}
r \to &\ r\,(1+X)^2\,e^{H\,t}, \\
& F(x) := \int \left((1-x^2)\,((1-x^2)^2 x^2 - 4\,K^2)  \right)^{-1}dx. \nn
\end{align}
This metric is the most simple nontrivial metric of the much studied class of spherically--symmetric metrics introduced by \CITE{McVittie, 1933}{33Vit}.
As already noted by \CITE{Robertson, 1927}{27Rob},
for $\Lambda=0$ it goes over to the Schwarzschild--metric in isotropic coordinates, whereas for $m=0$ the expanding de Sitter--metric results.
So we have effectively a cosmological model with Dark Energy containing a Schwarzschild Black Hole.

For such an explicitly time--dependent metric apparently (generalized) circular orbits seem not to have been analyzed yet.
In \CITE {Carrera and Giulini, 2010}{10CaG} an approximation is made for the McVittie Ansatz assuming small velocities,
leading to the time--development of the areal radius $R := {r\,(1+X)^2 e^{Ht}}$. 
In the case of the Schwarzschild--Robertson metric it effectively reduces to
\begin{equation}
\label{giul_eqn}
\ddot R = J^2/R^3 - M/R^2 + H^2\,R.
\end{equation}
In this case we can even define an effective potential 
\begin{equation}
V = M/R - \tfrac{1}{2}\, J^2/R^2 + \tfrac{1}{2}\,H^2 R^2.
\end{equation}
This admits circular orbits in the proper sense (among time--dependent solutions) by requiring $V=E=\const,\dot R = 0$.
Solving for $J^2$ this gives
\begin{equation}
J^2_{\scriptscriptstyle a\!p\!p\!r} = M\,R - H^2\,R^4,
\end{equation}
which makes sense in the range $0<R<(M/H^2)^{1/3} = K^{-2/3}M$.
The upper existence bound $R_x := K^{-2/3}M$ thus agrees exactly with the one obtained for the corresponding Schwarzschild--Kottler metric.
For the upper stability bound we get however $R_s := (2\,K)^{-2/3}M$, which is somewhat higher than the one for the Schwarzschild--Kottler metric.

In the case of the Schwarzschild--Robertson metric we could dispense with the above approximation and directly use the somewhat unwieldy coordinate--transformation of \teqns{rob_t}{rob_r} to get all quantities of interest based on the proper circular orbits of the Schwarzschild--Kottler metric.

The more conventional possibity would be to first find the timelike Killing--vector and then apply the standard approach based on the effective potential.

However, the J--based approach allows a much more direct derivation.
Recalling that the areal radius $R$ is an invariant of any spherically--symmetric metric,
and that $J^2$ is a scalar function, then from any solution $J^2(r)$ in terms of the standard radial coordinate $r:=e^c$, we can immediately read--off 
\begin{equation}
J^2(R) := J^2(r),
\end{equation}
using the same function $J^2$.
Also, from $\tdot r=0$ then follows $\tdot R=0$ (with respect to their corresponding affine parametrizations).
For the Schwarzschild--Robertson metric, and using the reduced areal radius $\varrho:=R/m$
with ${R = r\,(1+X)^2e^{Ht}}$,
\begin{equation}
J^2_{\scriptscriptstyle S\!R} = m^2 \varrho^2\,\frac{1-K^2\varrho^3}{\varrho-3},
\quad \text{where} \quad \varrho = y\left( 1 + \frac{1}{2\,y} \right)^2, \quad y:=\frac{r}{m}\,e^{Ht},
\end{equation}
with inverse function $\varUpsilon$ in the branch $\varrho\geq2$,
\begin{equation}
y = \varUpsilon(\varrho) := \tfrac{1}{2}\,( \varrho-1 ) + \tfrac{1}{2}\,\varrho\,( 1 - 2/\varrho )^{1/2}.
\end{equation}
Observe that for $y\to\infty$, $J^2_{\scriptscriptstyle S\!R} = J^2_{\scriptscriptstyle a\!p\!p\!r}$,
giving additional support to the above approximation, which was based on the 
assumption of small velocities.

A big advantage of our approach is that now from $J^2_{\scriptscriptstyle S\!R}$ we can calculate any other relevant quantity, like the tangent to the geodesics,~$T$.\footnote
{too complex to be fully displayed --- however the quotients $\dot {\bf J}/{\bf J}'$ simplify significantly}
Calculating it conventionally by means of a coordinate transformation from the SK--solution would have been relatively cumbersome.
Evidently, $T$ will now also get a $\p_r$--component.
As an example, we will show how the local tangential velocity gets modified
when chosing co--expanding observers $U=\frac{1+X}{1-X}\,\p_t$ (with expansion $\Theta = 3\,H$).
Analogously as in the previous derivation of \eqn{v2_sds},
\begin{align}
v^2_{\scriptscriptstyle S\!R} := &\ \frac{e^{2a}\tdot t^2-1}{e^{2a}\tdot t^2} =
\frac{{\bf J}/\varrho^2 - (e^{b-a}\,\dot {\bf J}/{\bf J}')^2}
{1+\bf J/\varrho^2}
= \frac{1-K^2\varrho^3 - Z^2}{\varrho-K^2\varrho^3-2}, \\
& \text{where} \quad {\bf J} := J^2_{\scriptscriptstyle S\!R}, \quad Z := \frac{e^{b-a}}{e^{K\tau}} \, \frac{\dot {\bf J}}{{\bf J}'} = K\,\frac{1}{\varUpsilon(\varrho)}\,\frac{\varUpsilon(\varrho)+2}{\varUpsilon(\varrho)-2}\,e^{K\tau},
\quad \tau := t/m. \nn 
\end{align}
Evidently, $v^2_{\scriptscriptstyle S\!R} < v^2_{\scriptscriptstyle S\!K}$. And if $v^2_{\scriptscriptstyle S\!R}>0$ at some time $\tau_0$, then for a fixed radius $\varrho$ there is a time $\bar\tau$, such that $v^2_{\scriptscriptstyle S\!R} \geq 0$ for $\tau_0\leq\tau\leq\bar\tau$ with $v^2_{\scriptscriptstyle S\!R}=0$ at $\tau=\bar\tau$ --- in this case the trajectory of the co--expanding observers has moved outside the existence region of the angular--momentum.
For small $K$ and $1 \ll \varrho$, as well as $\tau\ll 1/K$ (corresponding to $t \ll 1/H$, the Hubble--time),
\begin{equation}
v^2_{\scriptscriptstyle S\!R} \approx v^2_{\scriptscriptstyle S\!K} - 
\frac{K^2}{\varrho^3}\,e^{2\,K\tau} + o(K^4,\varrho^{-4},\tau^2).
\end{equation}
Thus the correction--factor to the Schwarzschild--Kottler velocity $v_{\scriptscriptstyle S\!K}$ of \eqn{v2_sk} for $\tau=0$ is $\Delta = -K^2/{\varrho^2}$.
For example, for a circular orbit at the rim of the Andromeda galaxy and $\tau=0$, the correction--factor is
$|\Delta| \approx 7.9\times10^{-26}$, and so would be completely negligible.

\subsection{Kerr--de Sitter Metric}
Here we will briefly apply our formalism to the equatorial orbits in Kerr--de Sitter spacetime.
Assuming time--independence of the metric,\footnote
{if $w'\neq0$, the Killing--vector $\p_t$ is not anymore hypersurface--orthogonal,
i.e. the metric is stationary only}
the J--\eqn{j_eqn_J} reduces to the quadratic
equation for $J^2/e^{2c}$,
\begin{equation}
\left( a' + J^2/e^{2c}(a'-c') \right)^2e^{2(a-c)} - w'^2\,J^2/e^{2c} \left( 1+ J^2/e^{2c} \right) = 0,
\end{equation}
with solutions 
\begin{equation}
\label{j2_kerr}
J^2/e^{2c} = \frac{a'\,(c'-a')\,e^{2(a-c)} + \tfrac{1}{2}\,w'^2 \pm \tfrac{1}{2}\,w'\! \left(4\,a'c'\,e^{2(a-c)} + w'^2 \right)^{1/2}} {(c'-a')^2\,e^{2(a-c)} - w'^2}.
\end{equation}

We base our derivations on the Kerr--de Sitter metric in the Boyer--Lindquist form as given by
\CITE {Stuchl\'ik and Slan\'y, 2004}{04StS}.\footnote
{there $y$ instead of $K^2$ is used and 
$a,r,t$ are in units of the mass~$m$}
With the auxi\-li\-ary quantities 
\begin{align}
\alpha := &\ \tfrac{a}{m}, \nn \\
x   := & \tfrac{m}{r}, \nn \\
K^2 := &\ \tfrac{1}{3}\,\Lambda\,m^2, \nn \\
L^2 := &\ K^2/x^2, \nn \\
M^2 := &\ 1 + K^2\alpha^2, \nn \\
N^2 := &\ 1+\alpha^2 x^2, \nn
\end{align}
the equatorial form of the metric derived from the full metric is defined by
\begin{align}
e^{2a} = &\ \frac{1}{M^4}\,\frac{\left(1-L^2 \right) N^2-2\,x }
{M^2N^2 + 2\,\alpha^2\,x^3 }, \\
e^{2b} = &\ \left(\left(1-L^2 \right) N^2-2\,x \right)^{-1}, \\
e^{2c} = &\ \frac{m^2}{M^4}\,x^{-2} \left(M^2N^2 + 2\,\alpha^2\,x^3 \right), \\
w = &\ -\frac{\alpha}{m}\,x^2 \,\frac{L^2N^2 + 2\,x}
{M^2N^2 + 2\,\alpha^2\,x^3}.
\end{align}
The expression for $J^2$ resulting from \eqn{j2_kerr} can be simplified to
\begin{align}
J^2_{\scriptscriptstyle K\!d\!S} = &\ \frac{m^2}{( 1 + {K}^{2}{\alpha}^{2} )^2}\,
\frac{\left(2\,\alpha+\alpha\,x\,(x^2+\alpha^2)\,K^2 
\mp x^{-1/2}(x^2+\alpha^2)\,\Delta_K^{1/2}  \right)^{2}}
{x^2\left( 1-3/x-\alpha^2K^2\pm 2\,\alpha\, x^{-3/2}\Delta_K^{1/2} \right)}, \\
&\ \text{where} \quad \Delta_K := 1 - K^2 x^3. \nn
\end{align}
Up to the factor $m^2( 1 + {K}^{2}{\alpha}^{2} )^{-2}$ this agrees with Stuchl\'ik and Slan\'y.\footnote
{they use an angular momentum $J$ rescaled by $(1 + {K}^{2}{\alpha}^{2})/m$}
And for $\alpha=0$ it agrees with the expression \eqn{jj_ksds} already derived for the Kottler--de Sitter metric, whereas for $K=0$ it agrees with the expression derived by \CITE {Bardeen, Press and Teukolsky, 1972}{72BPT},\footnote
{see also \CITE{Chandrasekhar 1983, ch. 7}{83Cha} for a more detailed derivation}
\begin{equation}
J^2_{\scriptscriptstyle K\!e\!r\!r} = m^2\,\frac{(x^2+\alpha^2\mp 2\,\alpha\,x^{1/2})^2}
{x\,(x^{2}-3\,x \pm 2\,\alpha\,x^{1/2})}.
\end{equation}

\subsection{Husain--Martinez--N\'u\~nez Metric}
Although in general an exact solution to the J--equation seems to be hopeless,
certain metrics admit a straightforward first--order approximate solution.

In the context of scalar field collapse some exact solutions to the Einstein field equation with a massless scalar have been found by \CITE{Husain--Martinez--N\'u\~nez, 1994}{94HMN}.\footnote
{see also \CITE {Faraoni 2015, ch. 4.8}{15Far} for a more detailed exposition}
Assuming without loss of generality $r>0$ and $t>-1/2H$ with $H\geq0$, the metric of their case I. can be more conveniently written as\footnote
{in fact they use a slightly different normalization, setting $m=1$ (their $c$)
and $at$ instead of our ``cosmological'' scale factor $1+2\,H\,t$, with $a=\pm1$ and $t$ with $at\geq0$}\footnote
{the sign of $\phi$ is independent of that of $\alpha$}
\begin{align}
\label{hmn_sol}
ds^2 = &\ (1+2\,H\,t)\left( -F^{\alpha}dt^2 + F^{-\alpha}dr^2 + r^2F^{1-\alpha}d\Omega^2 \right), \\
\phi = &\ \pm \frac{1}{6} \sqrt6\,\ln{(F^{\alpha}(1+2\,H\,t)^{3})}, 
\quad F := 1 - 2\,\frac{m}{r}, \quad \alpha := \pm\frac{1}{2} \sqrt{3}.
\end{align}
The constant $H$ can be interpreted as the Hubble--Parameter of the corresponding cosmological FLRW--metric (with scale--factor $a(t)=(1+3\,H\,t)^{1/3}$) when setting $m=0$. 
On the other hand, for $H=0$, we get a particular solution with para\-meter $\alpha = \pm\sqrt{3}/2$ and effective mass $M:=\alpha\,m$ of the two--parameter family of massless scalar solutions with $m\neq0$ and $\alpha^2\leq1$, well--known under the diverse names (in chronological order of their rediscoverers) of Fisher ('48), Buchdahl ('59), New\-man--Ja\-nis--Wi\-ni\-cour ('69) and Wyman ('81).\footnote
{also \CITE {Bronnikov, 1973}{73Bro} should be mentioned}\footnote
{i.e., up to isometries}
As already shown by \CITE{Buchdahl, 1959}{59Buc}\footnote
{in the following this seminal paper will be cited only by name}\footnote
{he uses the opposite signature of the metric and our parameter $\alpha$ is denoted by $\beta$} 
this family of solutions not only contains the Schwarzschild--solution as the special case $\alpha=1$ with $m>0$, but moreover is PPN--compatible with it,
as long as the effective mass $M$ is positive.
So the standard gravitational tests cannot fix the value of $\alpha$, except when going to the near--field, where $(m/r)^2$--terms become appreciable.
\smallskip

Let us first calculate the local circular velocity (with respect to static observers)
for the Buchdahl--metric. Our static J--algorithm immediately gives 
\begin{equation}
\label{v2_buch}
v^2_{\scriptscriptstyle B\!u\!c\!h} := \frac{a'}{c'} = \frac{\alpha\,m}{r - (1+\alpha)\,m} = 
\alpha\,\frac{m}{r}\left(1 + (1+\alpha)\,\frac{m}{r} + o\left(\frac{m}{r}\right)^2 \right),
\end{equation}
which in fact up to $o\left({m}/{r}\right)^2$--terms seems to coincide with the circular velocity corresponding to an effective Schwarzschild--mass $M=\alpha\,m$. However, for the proper existence of the velocity (i.e. bound circular orbits), we must in addition assume $M>0$.
This restricts the general HMN--solution to solutions where the ``naked mass'' $m$ has the same sign as the parameter $\alpha = \pm\sqrt{3}/2$.
In fact, as shown by Buchdahl, there is a discrete isometry $m\to-m,\:\alpha\to-\alpha$ between the apparently different metrics. Therefore, we will restrict in what follows to the ``physical metric'' with parameters $\alpha=+\sqrt{3}/2$ and $m>0$.
\smallskip

Now to the approximate solution for the HMN--metric.
Noting that this metric is conformally related with factor $1+2Ht$ to the above static Buchdahl--metric, this suggests the following {\em Ansatz} 
\begin{equation}
J^2_{\scriptscriptstyle H\!M\!N} := \frac{a'}{c'-a'}\,e^{2c}(1+2Ht), 
\end{equation}
where the functions $a$ and $c$ are taken from the corresponding static Buchdahl--metric with $M>0$.
Inserting this Ansatz into the J--\eqn{j_eqn_J} shows that it is satisfied up to $o(H^2)$--terms --- more precisely up to $1/m\,o(K^2x)$, where again $K:=Hm$ and $x:=r/m$. 
In terms of the areal radius $\varrho := r\,(1+X)^2 (1+2Ht)^{1/2}$,
\begin{equation}
J^2_{\scriptscriptstyle H\!M\!N} \approx {M}{\varrho}\,(1+Ht), \quad M := \alpha\,m,
\end{equation}
for $\varrho\to\infty$ and up to $o(H^2)$.
\smallskip

Taking again the Andromeda galaxy as an example, the approximate distance where our approximation begins to break down is $d \approx 7.5\times10^{21}\:ly$.
This is many orders of magnitude higher than the extent of our observable universe estimated to be $r \approx 9.3\times10^9\:ly$ (slightly below the Hubble--radius $r_H:=H^{-1}\approx1.3\times10^{10}\:ly$).
And one of the biggest known cosmological structures, the local Laniakea supercluster, has an radial extent of $r \approx 2.6\times10^8\:ly$, with about $10^5$ galaxies ($m \approx 3.9\times10^9\:ly$).
The breakdown radius is reduced to $d \approx 4.5\times10^{15}\:ly$, which is still significantly higher than the extent of our observable universe (even when allowing a mass--factor of 10 for interstellar gas and radiation).
Therefore for all practical purposes, even for the biggest known structures, for this particular cosmological model based on the HMN--solution, we can safely ignore the $H^2$-- and any higher--order terms.

Also the velocity $v^2$ with respect to a coexpanding observer gets the same Hubble--factor (1+Ht).
This can be seen as follows. 
First we get, up to ignorable $K^2x$--terms, using \eqn{t_def_J},
\begin{equation}
-g_{tt}\tdot t^2 = P^2 \equiv 1 + J^2_{\scriptscriptstyle H\!M\!N} e^{-2c}/(1+2Ht)
\equiv \frac{a'}{c'-a'}\,e^{2c},
\end{equation}
as in the static case of \eqn{j2_eqn}.
Then, up to ignorable $K^2x$--terms,
\begin{equation}
v^2_{\scriptscriptstyle H\!M\!N} = \frac{\alpha\,m}{r - (1+\alpha)\,m}.
\end{equation}
This is {\em exactly} the expression of the velocity of \eqn{v2_buch} for the corresponding static Buchdahl--metric and apparently time--independent. However, when again expressed in terms of the areal radius,
\begin{equation}
v^2_{\scriptscriptstyle H\!M\!N} \approx \frac{M}{\varrho}\,(1+Ht).
\end{equation}

Of course, this HMN--metric is very special in that it is not only conformally static, 
but also in that the squared cosmological scale--factor is linear in time.
A slight generalization where our approximation still works would consist in taking an arbitrary power of the above scale--factor --- however the corresponding field equations would not anymore be satisfied. Therefore the validity of our approximation referring to the HMN--metric must be considered as fortuitious.\footnote
{in particular this approximation fails for the solution of Sultana--Dyer (in the conformally Schwarzschild form of the metric as given by \CITE {Faraoni 2015, ch. 47}{15Far})}~Trying to conventionally solve the relatively complex geodesic equation, this approximate solution would hardly have been found. This again shows the superiority of the J--based approach.

\section{Discussion}
\label{disc_sec}
Let us emphasize again that we could of course have done the static analysis in section~\ref{static_sec} using the conventional approach based on an effective potential. However here we calculated the relevant quantities in a more transparent and shorter way using only $J^2$ and its first derivative. In fact, the usefulness of the angular--momentum function seems not to have been recognized so far. 
However, the full potential of our approach would only be revealed when analyzing effectively time--dependent metrics, like inhomogeneous isotropic cosmologies.
In particular, our approach would allow to find out if the asymptotically flat galactic rotation curves could be understood as a manifestation of some dynamic gravitational effects --- without invoking either dark matter or Milgrom's phenomenological MOND--theory.
Unfortunately such cosmological models cannot be analyzed ana\-ly\-ti\-cally, except when exact solutions to the {J--equation} can be found, which is highly improbable.
Nevertheless, appropriate approximate solutions to the J--equation could be constructed.

However we were able to analyze two particular time--dependent spacetimes: \\
i) \hspace{1.85pt}a metric with $\Lambda$--term 
(the Schwarzschild--Robertson solution) and \\
ii) a metric with Hubble--expansion
(the Husain--Martinez--N\'u\~nez solution). \\
Whereas in the first example cosmological effects first appear in $o(\Lambda)=o(H^2)$
and turn out to be negligible for galaxies, in the second example both $J^2$ and $v^2$ are in $o(H)$ subject to the local Hubble--flow with scale--factor $a=1+Ht$.

Perhaps more realistic such cosmological models will be needed,
as well as generally applicable approximation methods in order to better understand and perhaps resolve the as yet empirical baryonic Tully--Fisher relation. \\
The present work can be seen as a step in this direction.

\section*{Acknowledgment}
I thank Peter Aichelburg for helpful discussions.


\begin{thebibliography}{999}

\bibitem {18Kot} F Kottler, 1918: \\
\tit{\"Uber die physikalischen Grundlagen 
der Einsteinschen Gravitationstheorie} 
\emph{Ann der Phys} \textbf{56}, 401 -- 462

\bibitem {27Rob} HP Robertson, 1927 : \\
\tit{Dynamical space--times which contain a conformal euclidean 3--space}   
\emph{Trans Am Math Soc} \textbf{29}, 481 -- 496

\bibitem {28Rob} HP Robertson, 1928 : \\
\tit{On Relativistic Cosmology}
\emph{Philos Mag} \textbf{5}, 835 -- 848

\bibitem {33Vit} GC McVittie, 1933 : \\
\tit{The Mass--Particle in an Expanding Universe}
\emph{MNRAS} \textbf{93}, 325 -- 339

\bibitem {59Buc} HA Buchdahl, 1959 : \\ 
\tit{Reciprocal Static Metrics and Scalar Fields\\
in the General Theory of Relativity}
\emph{Phys Rev} {\bf 115}, 1325 -- 1328

\bibitem {72BPT} JM Bardeen, WH Press, SA Teukolsky, 1972 : \\
\tit{Rotating Black Holes: Locally Nonrotating Frames, \\
Energy Extraction, and Scalar Synchrotron Radiation}
\emph{ApJ} {\bf 178}, 347 -- 369

\bibitem {73Bro} KA Bronnikov, 1973 : \\
\tit{Scalar--Tensor Theory and Scalar Charge}
\emph{Acta Phys Polon} \textbf{B4}, 251 -- 266

\bibitem {83Cha} S Chandrasekhar, 1983 : \\
{\em The Mathematical Theory of Black Holes} \\
Clarendon Press (Oxford)

\bibitem {84Wal} RM Wald, 1984 : \\
{\em General Relativity}, \\
The University of Chicago Press (Chicago)

\bibitem {94HMN} V Husain, EA Martinez, D N\'u\~nez, 1994 : \\
\tit{Exact solution for scalar field collapse}
\emph{Phys Rev D} \textbf{50}, 3783 -- 3786
\\ \url{http://arXiv.org/pdf/gr-qc/9402021}

\bibitem {96Hay} SA Hayward, 1996 : \\
\tit{Gravitational energy in spherical symmetry}
\emph{Phys Rev D} \textbf{53}, 1938 -- 1949
\\ \url{http://arXiv.org/pdf/gr-qc/9408002}

\bibitem {98Fra} T Frankel, 1998 : \\
{\em The Geometry of Physics}, \\
Cambridge University Press (Cambridge)

\bibitem{99StH} Z Stuchl\'ik, S Hled\'ik, 1999 : \\
\tit{Some properties of the Schwarzschild--de Sitter \\
and Schwarzschild--anti--de Sitter spacetimes}
\emph{Phys Rev D} \textbf{60}, 044006

\bibitem {03KHM} AW Kerr, JC Hauck, B Mashhoon, 2003 : \\
\tit{Standard Clocks, Orbital Precession and the Cosmological Constant }
\emph{Class Quant Grav} \textbf{20}, 2727 \\
\url{http://arXiv.org/pdf/gr-qc/0301057}

\bibitem {04Car} SM Carroll, 2004 : \\
{\em Spacetime and Geometry}, \\
Addison Wesley (San Francisco)

\bibitem {04Rob} MD Roberts, 2004 : \\
\tit{Galactic Metrics}
\emph{Gen Rel Grav} \textbf{36}, 1 -- 10
\\ \url{http://arXiv.org/pdf/astro-ph/0209456}

\bibitem {04Lak} K Lake, 2004 : \\
\tit{Galactic Potentials}
\emph{Phys Rev Lett} \textbf{92}, 051101 \\
\url{http://arXiv.org/pdf/gr-qc/0302067}

\bibitem {04StS} Z Stuchl\'ik, P Slan\'y, 2004 : \\
\tit{Equatorial circular orbits in the Kerr--de Sitter spacetimes}
\emph{Phys Rev D} \textbf{69}, 064001 \\
\url{http://arXiv.org/pdf/gr-qc/0307049}

\bibitem {05SuD} J Sultana, CC Dyer, 2005 : \\
\tit{Cosmological black holes: A black hole 
in the Einstein--de Sitter universe}
\emph{Gen Rel Grav} \textbf{37(8)}, 1349 -- 1370

\bibitem {05BeH} R Beig, JM Heinzle, 2005 : \\
\tit{CMC--Slicings of Kottler--Schwarzschild--de Sitter Cosmologies} 
\emph{Comm Math Phys} \textbf{260}, 673 -- 709 \\
\url{http://arXiv.org/pdf/gr-qc/0501020}

\bibitem {05Gau} SS McGaugh, 2005 : \\
\tit{The Baryonic Tully--Fisher Relation of Galaxies \\
with Extended Rotation Curves 
and the Stellar Mass of Rotating Galaxies}
\emph{APJ} \textbf{632}, 859 -- 871 \\
\url{http://arXiv.org/pdf/astro-ph/0506750}

\bibitem {06ONe} B O'Neill, 2006 : \\
{\em Semi--Riemannian Geometry with Applications \\
to General Relativity}, \\
Academic Press (New York)

\bibitem {07FaJ} V Faraoni, A Jacques, 2007 : \\
\tit{Cosmological expansion and local physics}
\emph{Phys Rev D} \textbf{76}, 063510 \\
\url{http://arxiv.org/pdf/gr-qc/0707.1350}

\bibitem {08Mil} M Milgrom, 2008 : \\
\tit{The MOND paradigm}
\emph{Talk presented at the XIX Rencontres de Blois ``Matter and energy in the Universe: from 
nucleosynthesis to cosmology'', May 2007}
\\ \url{http://arXiv.org/pdf/0801.3133}

\bibitem {10CaG} M Carrera, D Giulini, 2010 : \\
\tit{Influence of global cosmological expansion \\
on local dynamics and kinematics}
\emph{Rev Mod Phys} \textbf{82}, 169 -- 208 \\
\url{http://arXiv.org/pdf/0810.2712}

\bibitem {11FrZ} VP Frolov, A Zelnikov, 2011 : \\
{\em Introduction to Black Hole Physics}, \\
Oxford University Press (Oxford)

\bibitem {12TTK} SG Turyshev, VT Toth, G Kinsella, SC Lee, SM Lok, J Ellis, 2012 : \\
\tit{Support for the thermal origin of the Pioneer anomaly}
\emph{Phys Rev Lett} \textbf{107(8)}, 81103 \\
\url{http://arXiv.org/pdf/1204.2507}

\bibitem {12HJA} X Hernandez, MA Jimenez, C Allen, 2012 : \\
\tit{Wide binaries as a critical test of Classical Gravity}
\emph{EPJC C} \textbf{72}, id 1884  \\
\url{http://arXiv.org/pdf/arXiv:1105.1873}

\bibitem {13KlP} PE Kloeden, C P\"otzsche, 2013 : \\
{\em Nonautonomous Dynamical Systems in the Life Sciences} \\
Springer International

\bibitem {14Nol} BC Nolan, 2014 : \\
\tit{Particle and photon orbits in McVittie spacetimes}
\emph{Class Quant Grav} \textbf{31}, 235008 \\
\url{http://arXiv.org/pdf/1408.0044}

\bibitem {15Far} V Faraoni, 2015 : \\
\tit{Cosmological and Black Hole Apparent Horizons }
Lecture Notes in Physics, Vol. 907 \\
Springer (New York/Berlin/Heidelberg)

\bibitem {16Abr} MA Abramowicz, 2016 : \\
\tit{Velocity, acceleration and gravity in Einstein’s relativity}
\url{http://arXiv.org/pdf/1608.07136}


\end{thebibliography}
\end{document}